\documentclass[aps,prl,showpacs,twocolumn,floatfix,superscriptaddress]{revtex4}
\usepackage{graphicx}

\begin{document}

\title{Non-equilibrium clustering of self-propelled rods}

\date{\today}

\author{Fernando Peruani}
\affiliation{Max Planck Institute for
Physics of Complex Systems, N\"othnitzer Str. 38, 01187 Dresden,
Germany} \affiliation{ZIH, TU Dresden, Zellescher Weg 12, 01069
Dresden, Germany}

\author{Andreas Deutsch}
\affiliation{ZIH, TU Dresden, Zellescher Weg 12, 01069 Dresden,
Germany}

\author{Markus B\"ar}
\affiliation{Physikalisch-Technische Bundesanstalt, Abbestr. 2-12,
10587 Berlin, Germany}

\begin{abstract}

Motivated by aggregation phenomena in gliding bacteria, we study
collective motion in a two-dimensional model of active, 
self-propelled rods interacting through volume exclusion.
In simulations with individual particles, we find that particle clustering is
facilitated by a sufficiently large packing
fraction $\eta$  or length-to-width ratio $\kappa$.
The transition to clustering in simulations is well captured by a
mean-field model for the cluster size distribution, which predicts 
that the transition values
$\kappa_c$ of the aspect ratio for a fixed packing fraction $\eta$
is given by $\kappa_c = C/\eta -1 $ where $C$ is a constant.

\end{abstract}


\maketitle


{\it Introduction.---} Emergent large-scale patterns of interacting
self-driven motile elements are observed in a wide range of
biological systems of different complexity: from human crowds,
herds, bird flocks, and fish schools \cite{reviews} 
to multicellular aggregates, e.g. of bacteria and
amoebae \cite{benjacob-aip} as well as sperms \cite{Riedel-Sperm}.
A recurrent question is how these entities
coordinate their behavior to form groups which move collectively.
At a theoretical level, several qualitative approaches have been made to
incorporate the diverse collective behaviors of such different
systems in a common framework
\cite{reviews,theocollectivemotion,active-brownian}.
More specific  models for bacteria like {\it E. coli}
as well as for amoebae like {\it D. discoideum}
\cite{benjacob-aip}, have been based on chemotaxis, a long-range cell interaction
mechanism according to which individual cells move in response to chemical signals produced
by all other cells.
However, in some bacteria there is no evidence
for chemotactic cues and cells coordinate their movement by
cell-to-cell signalling mechanisms in which physical contact
between bacteria is needed  \cite{myxoreview}.
Consequently, one may ask how such bacteria aggregate in order to
communicate.

Another relevant aspect is the influence of the shape of the bacteria.
The shape  has been shown to be essential for
individual motion of swimming bacteria \cite{sizebacteria}.
In contrast, the role of the cell shape for collective motion has
remained mostly unexplored.
It has been demonstrated experimentally \cite{gruler}
that migrating elongated amoeboid cells  exhibit alignment effects similar to
those reported in liquid crystals \cite{onsager}. 
A prominent example for collective behavior 
with no apparent long range interactions are the striking patterns
 observed during the life-cycle of gliding myxobacteria, see e.g.
\cite{myxoreview, myxoexperiments}.
Earlier modeling work has reproduced many of these
patterns in three dimensions assuming either perfect alignment \cite{boerner-prl}
or a phenomenological alignment force \cite{myxomodels}.
These models have all considered patterns resulting from exchange of chemical signals, 
that are absent in an early stage of the myxobacterial life cycle.
Nevertheless, a trend from initial independent motion
towards formation of larger clusters of aligned bacteria is often observed ( Fig. 1).

Here, we study a model of self-propelled rods that have only
repulsive excluded volume interactions in two dimensions.
We find that the interplay of  rod geometry, self-propulsion
and  repulsive short-range interaction is sufficient to facilitate
aggregation into clusters.
%
\begin{figure}
\centering
\includegraphics[scale=0.4]{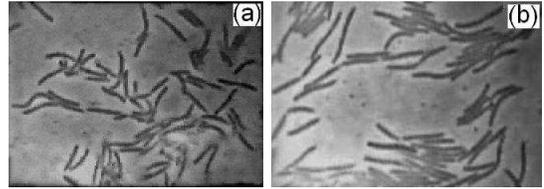}
\caption{Example for clustering of myxobacteria ({\it M. xanthus}) in the
early stage of the life cycle. (a) Immediately after maturation of
spores. (b) Afterwards, during the vegetative phase. Snapshots are taken
from a movie (Ref. \cite{myxoexperiments}(b)), the frame size is 
40 x 30 $\mu m^2$. Similar phenomena 
were seen in other bacterial species ({\it cf.} Ref. \cite{myxoreview}(a)). }
\label{experiment}
\end{figure}
%
\begin{figure}
\centering
%
{\includegraphics[width=0.32\linewidth,
height=0.32\linewidth]{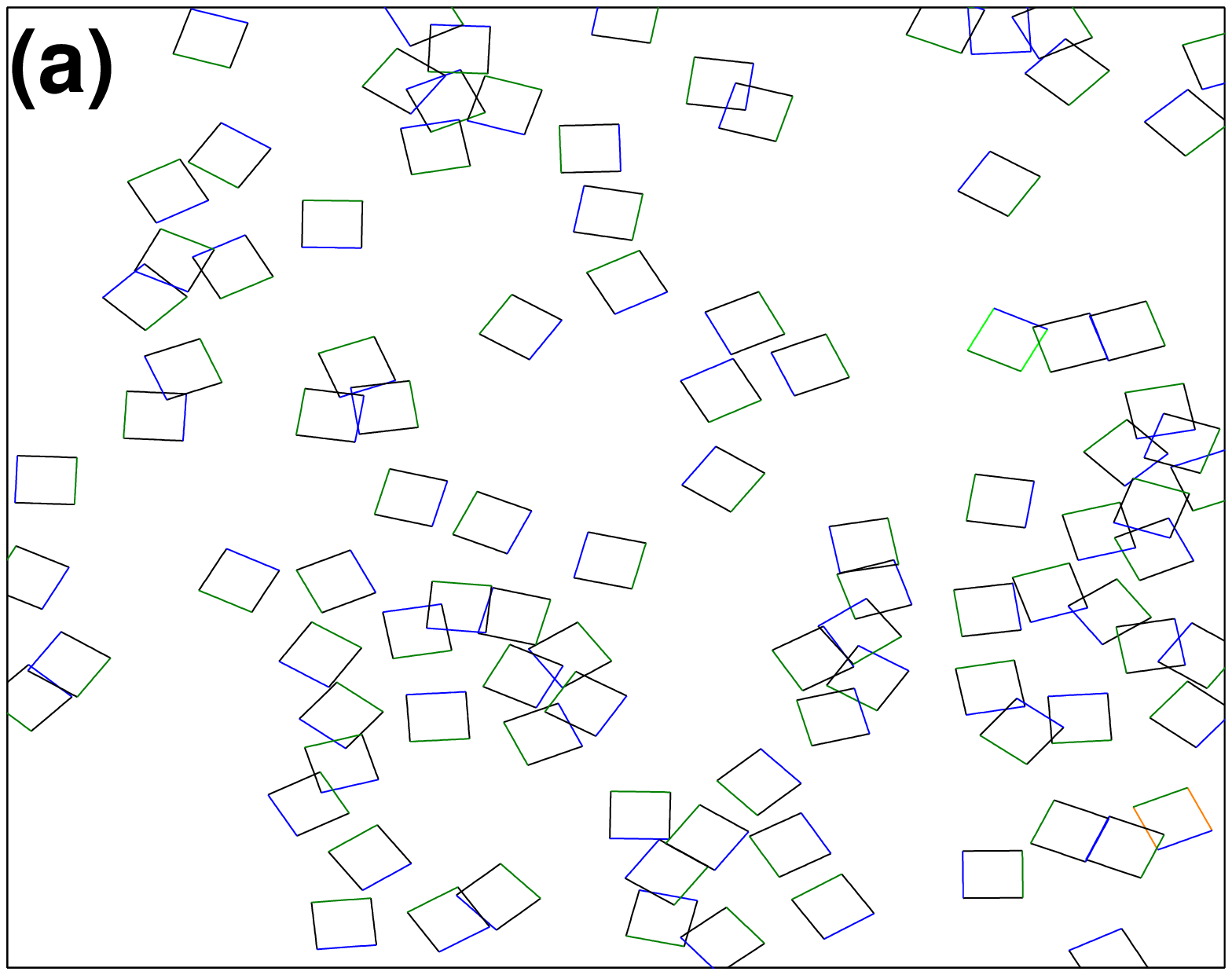}
\includegraphics[width=0.32\linewidth,
height=0.32\linewidth]{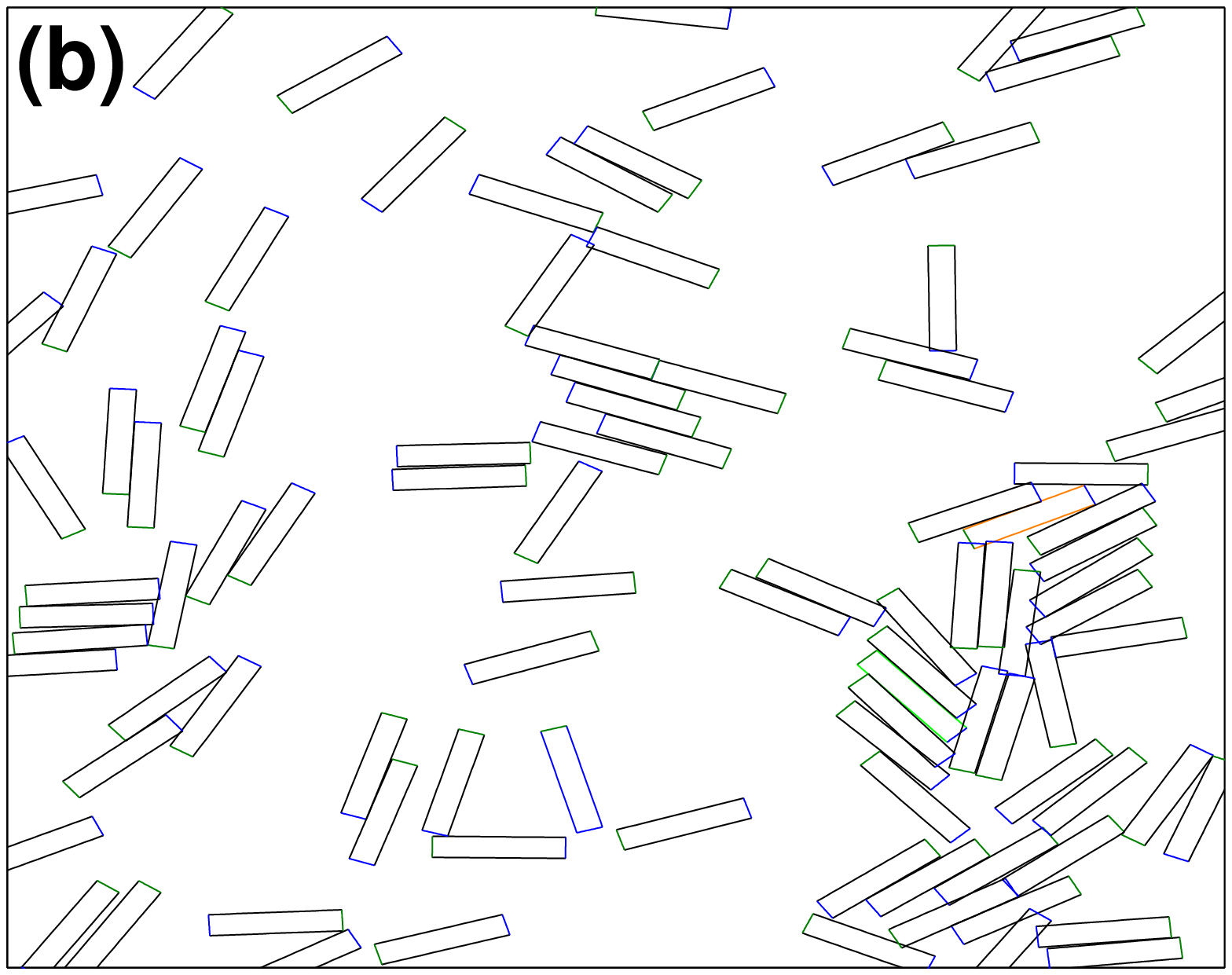}
\includegraphics[width=0.32\linewidth,
height=0.32\linewidth]{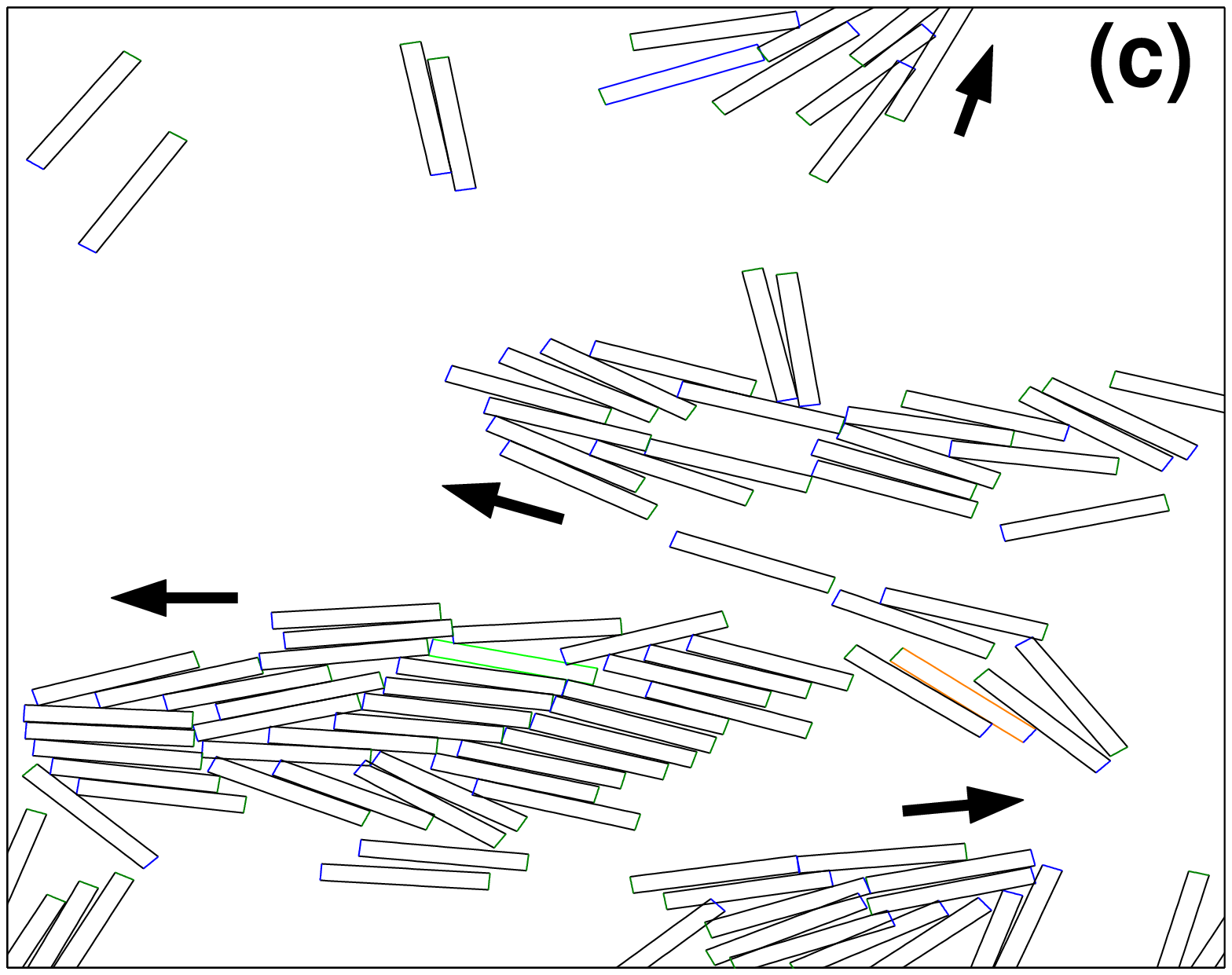}}
%
{\includegraphics[width=0.32\linewidth,
height=0.32\linewidth]{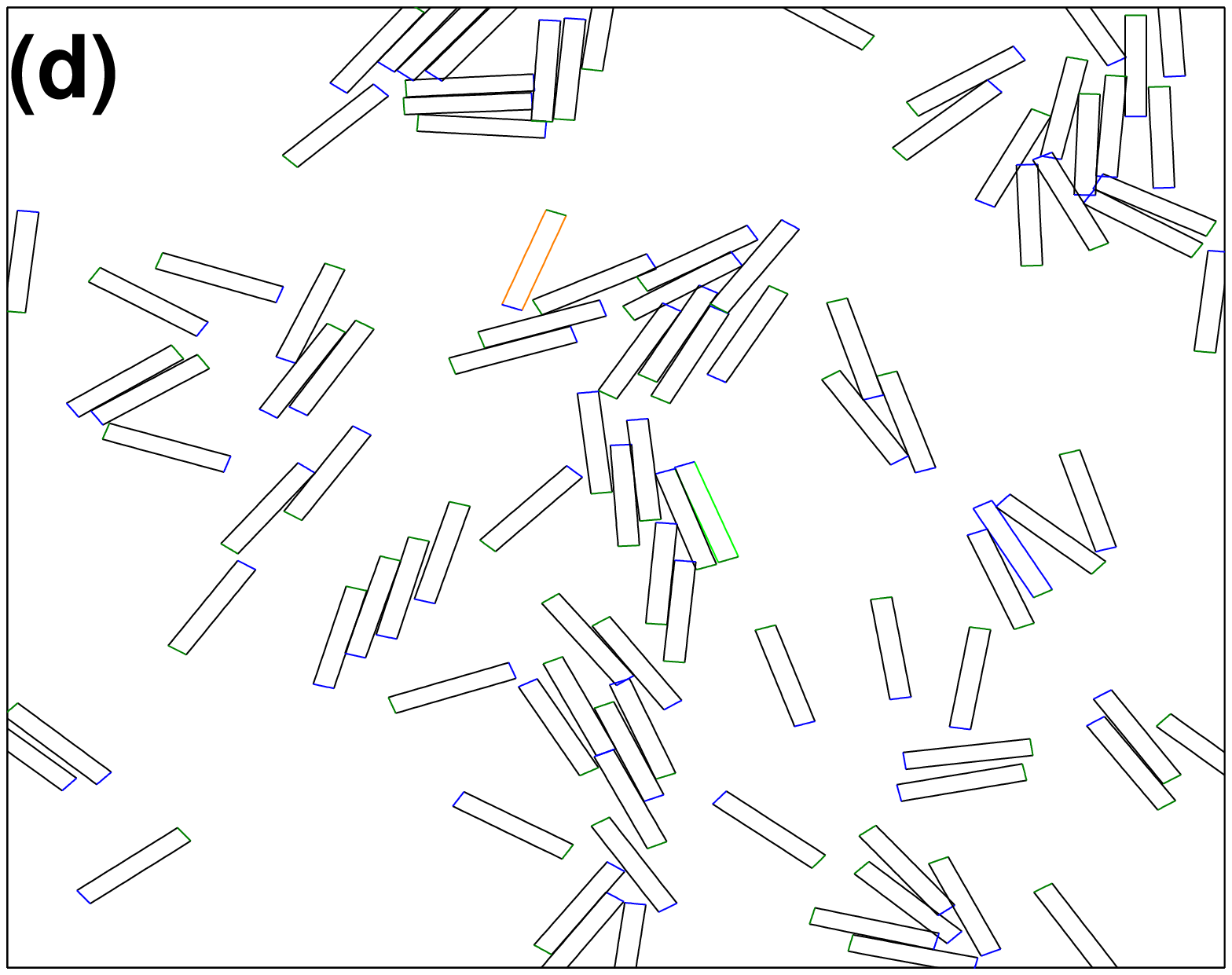}
\includegraphics[width=0.32\linewidth,
height=0.32\linewidth]{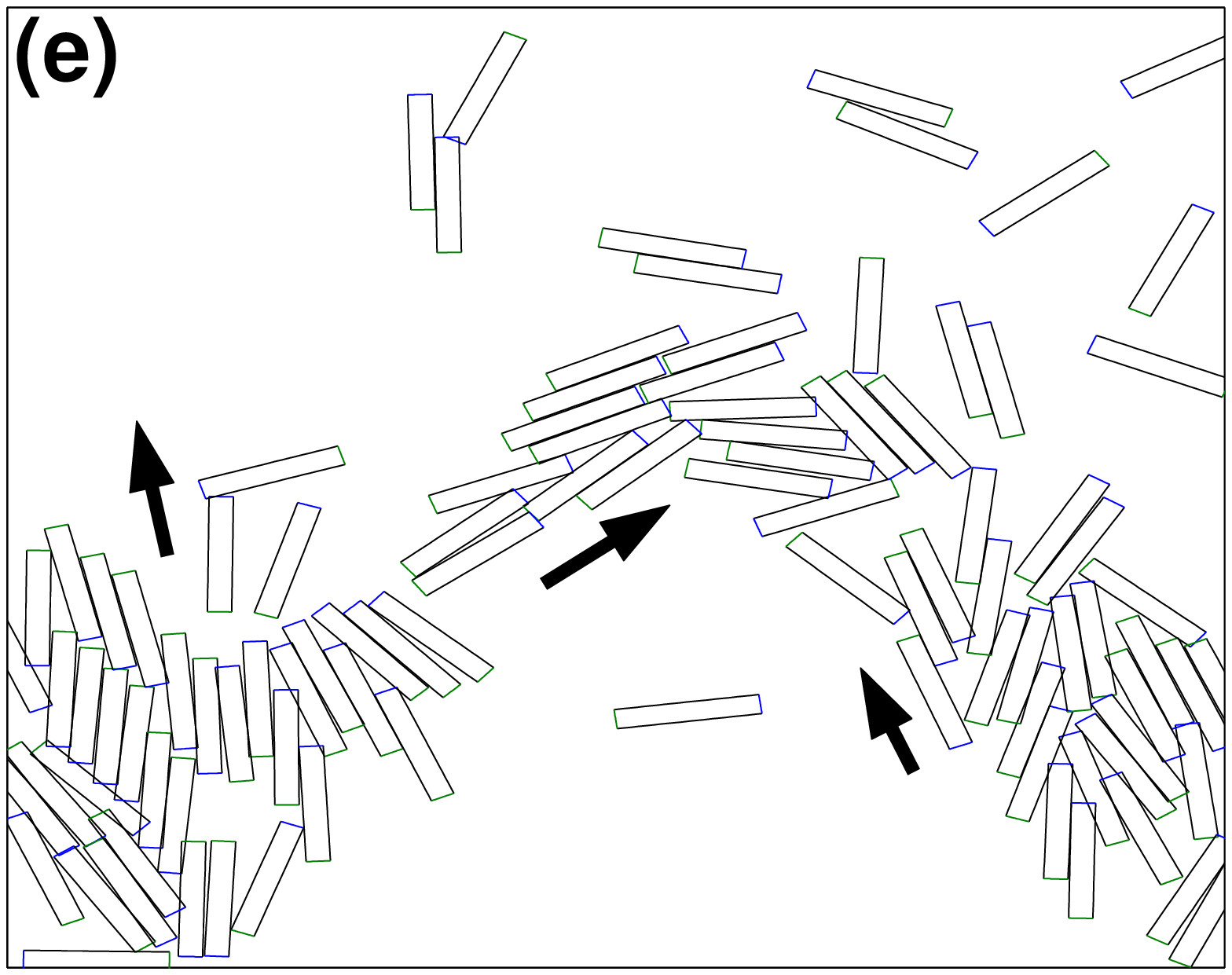}
\includegraphics[width=0.32\linewidth,
height=0.32\linewidth]{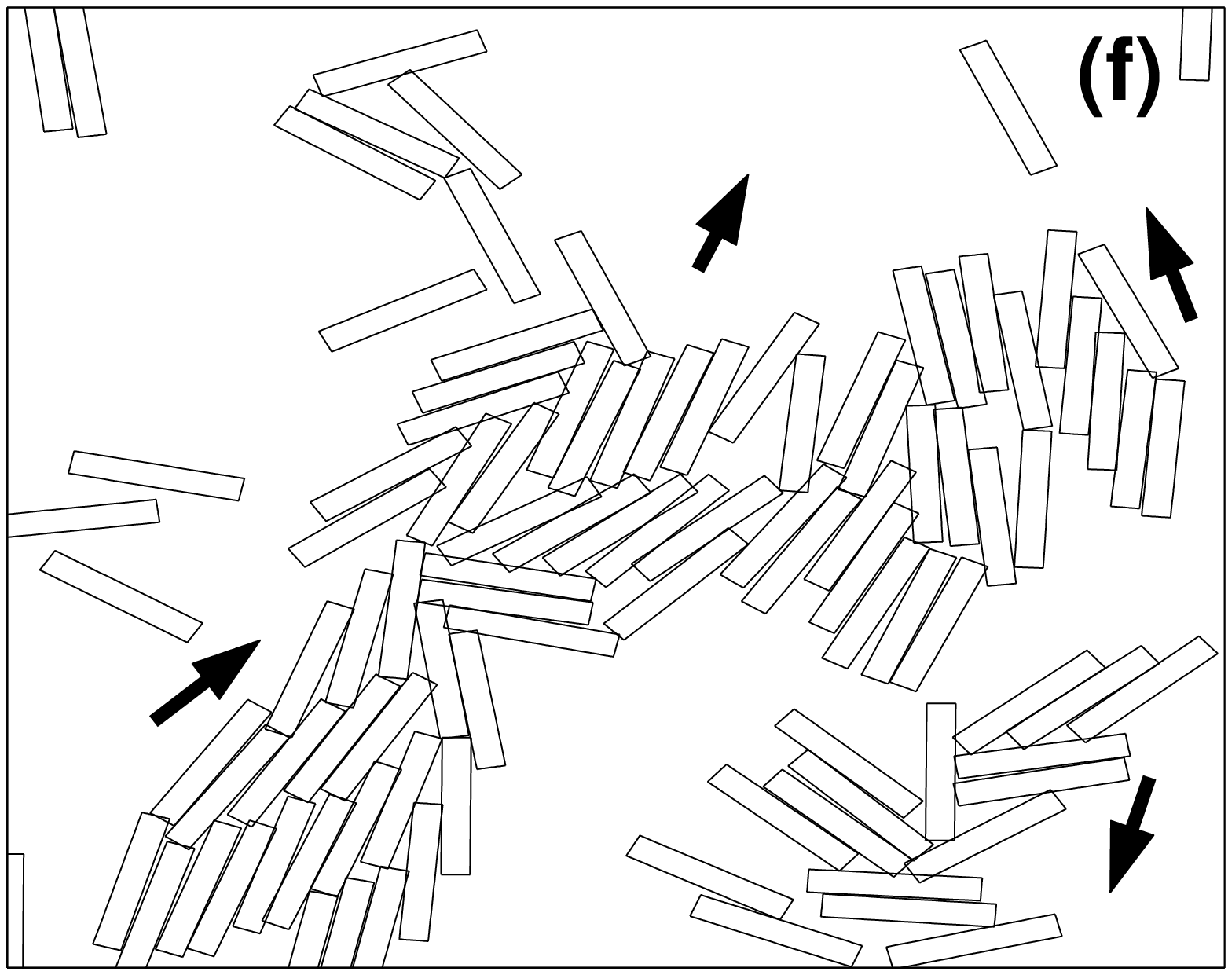}}
\caption{Simulation snapshots of the steady states for different
particle anisotropy $\kappa$ and the same packing fraction $\eta$
(a-c), and the same $\kappa$ and different $\eta$ (d-f). Fixing
 $\eta=0.24$: (a) before the transition, $\kappa=1$; (b) almost at
the transition, $\kappa=5$; (c) after the transition, $\kappa=8$.
Fixing $\kappa=6$: (d) before the transition, $\eta=0.18$; (e)
just crossing the transition, $\eta=0.24$; (f) after the
transition, $\eta=0.34$. In all cases, particles $N=100$ and
particle area $a=0.2$. The arrows indicate the direction of motion
of some of the clusters.} \label{fig2}
\end{figure}
In simulations of an individual based model (IBM), clustering of
self propelled particles (SPP) is observed for large enough
packing fraction $\eta$ resp. aspect ratio  $\kappa$ of the rods
(see Fig. \ref{fig2}).
We define the onset of clustering by the transition from a unimodal
to a bimodal cluster size distribution.
A mean-field approximation (MFA) for the cluster size distribution is derived and
reproduces the change from a unimodal to a bimodal shape upon
increase of either $\eta$ or $\kappa$.
The MFA yields a simple equation $\kappa_c = C/\eta - 1$
for the critical rod aspect ratio, $\kappa_c$, at the onset of clustering in line 
with the IBM simulation results. 
%
fitted with $C = 1.46$.
If diffusion is added to the active motion (active Brownian rods), 
the clustering transition is shifted to higher values of $\kappa$, 
whereas clustering is absent for pure diffusive motion (Brownian rods)
as well as for isotropic particles with $\kappa = 1$. 
Hence, clustering of particles with excluded volume interaction
requires both active motion, i. e. a non-equilibrium system, and
elongated particles (= rods).

{\it Individual-based model (IBM).---}
 We consider $N$ rod-like particles moving on a
plane. 
Each particle is equipped with a self-propelling force acting
along the long axis of the particle. 
We assume that particles are submerged in a viscous medium. 
Velocity and angular velocity are proportional to the force and torque,
correspondingly.
The rod-shape of the particles requires 
three different friction coefficients which correspond to the
resistance exerted by the medium when particles either rotate or move along
their long and short axes.
Inertial terms are neglected (overdamped motion). 
Consequently the movement of the i-th rod is governed
by the following equations for the velocity of its center of
mass and angular velocity:
\begin{eqnarray} \label{update_position}
(v^{(i)}_{\parallel}, v^{(i)}_{\perp}) &=&  \left(
\frac{1}{\zeta_{\parallel}}(F-\frac{\partial U^{(i)}}{\partial
x_{\parallel}}),
 -\frac{1}{\zeta_{\perp}} \frac{\partial U^{(i)}}{\partial x_{\perp}} \right) \nonumber \\
\label{update_orientation} \dot{\theta}^{(i)} &=&
-\frac{1}{\zeta_{\theta}}\frac{\partial U^{(i)}}{\partial \theta}
\end{eqnarray}
where $v^{(i)}_{\parallel}, v^{(i)}_{\perp}$ refer to the
velocities along the long and short axis of the rods,
respectively,
 $\zeta_{i}$ indicates the corresponding friction
coefficients ($\zeta_{\theta}$ is related to the friction torque),
$U^{(i)}$ refers to the energy of the interaction of the i-th rod with all other rods, 
and $F$ is the magnitude of the self-propelling force. 
The motion of the center of mass $\mathbf{\dot{x}}^{(i)}= (v_x^{(i)},v_y^{(i)})$ 
of the i-th rod is given by 
\begin{eqnarray} 
v_x^{(i)} = v^{(i)}_{\parallel} \cos \theta^{(i)} +  v^{(i)}_{\perp} \sin \theta ^{(i)} \nonumber \\
v_y^{(i)} = v^{(i)}_{\parallel} \sin \theta^{(i)} -  v^{(i)}_{\perp} \cos \theta ^{(i)} 
\end{eqnarray}
Particles interact by ,,soft´´ volume exclusion, {\it i. e. } by 
a potential that penalizes particle overlaps in the following way:
\begin{eqnarray} \label{potential}
U^{(i)}(\mathbf{x}^{(i)},\theta^{(i)},\mathbf{x} ^{(j)},\theta^{(j)})& & \nonumber \\ 
= \phi \sum_{j=1,j \neq i}^{N} 
 \left( (\gamma - 
a_{o}(\mathbf{x}^{(i)},\theta^{(i)},\mathbf{x}^{(j)},\theta^{(j)} ))^{- \beta} -
\gamma^{- \beta} \right) & & 
\end{eqnarray}
where $a_{o}(\mathbf{x}^{(i)},\theta^{(i)},\mathbf{x}^{(j)},\theta^{(j)})$ 
is the area overlap of the rods $i$ and $j$,
$\gamma$ is a parameter which can be associated to the maximum
compressibility, $\beta$ controls the stiffness of the particle, 
and $\phi$ is the interaction strength.
The simulations were performed placing $N$ identical particles
initially at random inside a box of area $A$ with periodic
boundary conditions. 
The values of the parameters are given in
\cite{parameters}.

\begin{figure}
\centering
\includegraphics[scale=0.3]{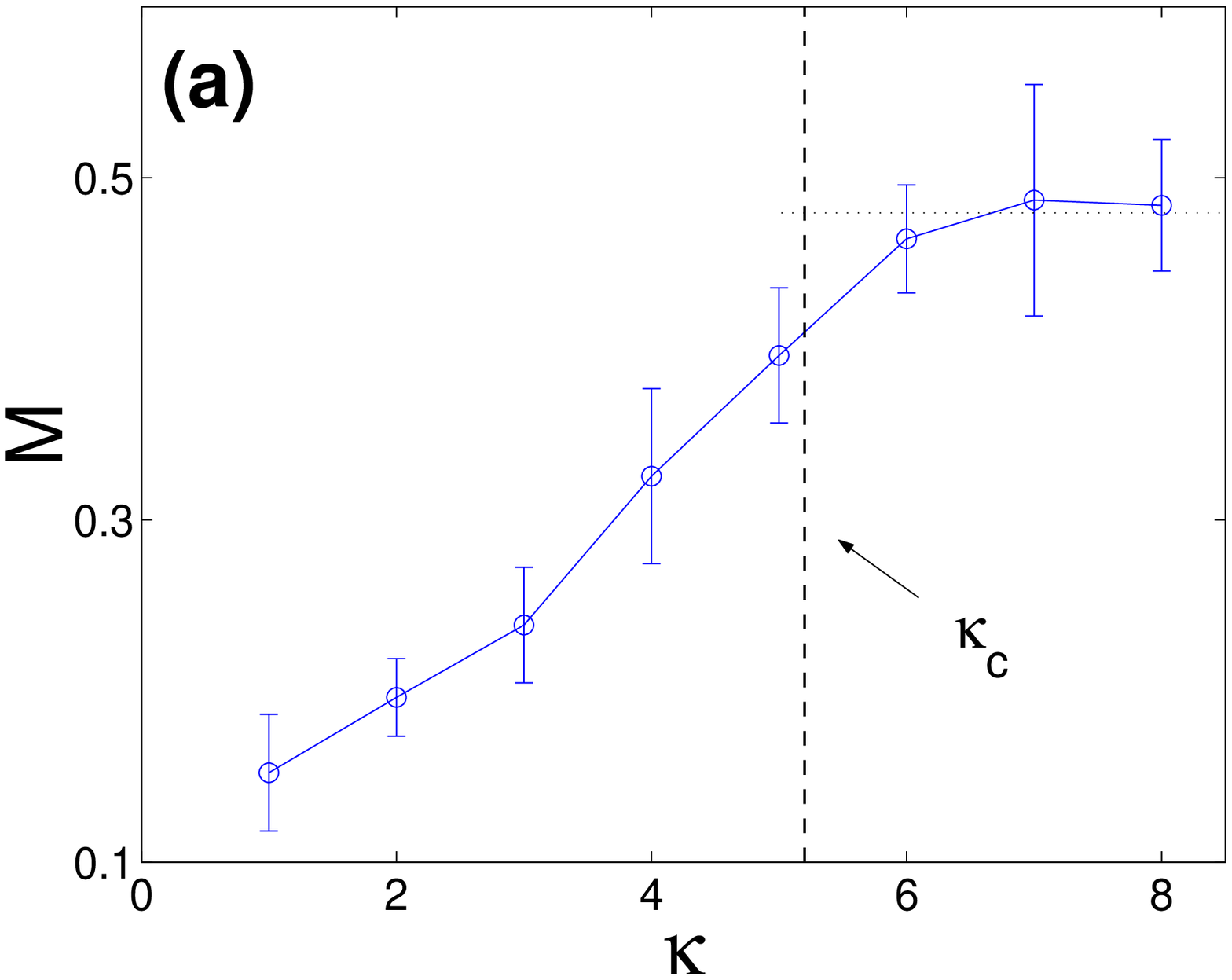}
\includegraphics[scale=0.3]{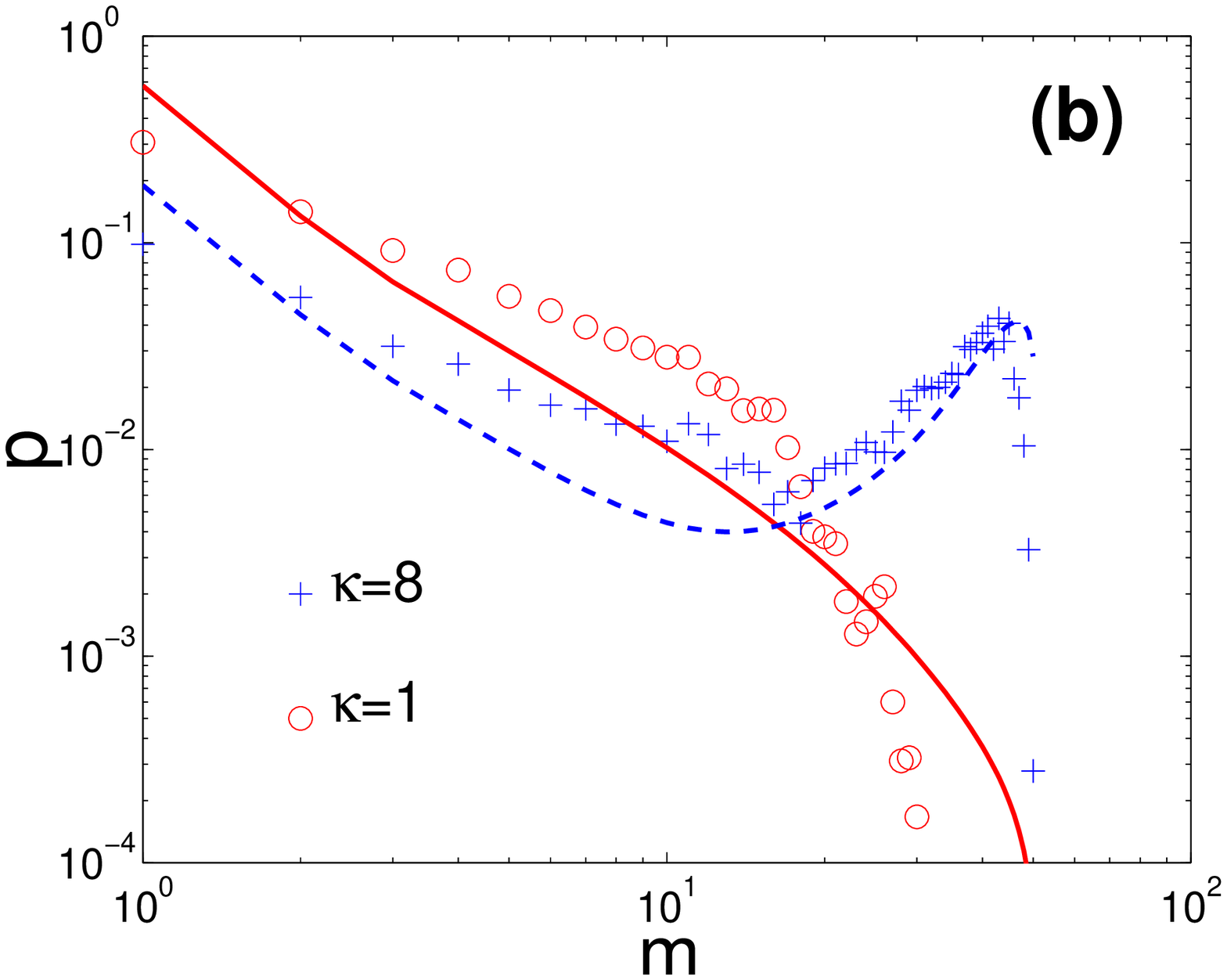}
\caption{(a) The mean maximum cluster size $M$ vs $\kappa$ for IBM
simulations ($N=50$). (b)  $p(m)$ as function of the cluster size $m$ for
$\eta=0.34$. Symbols
show the average over eight IBM simulations for active particles with $N=50$ and
$\kappa=1$ (circles) and $\kappa=8$ (crosses), errors bases give distributions
of individual runs. The lines
correspond to the mean field theory for $\kappa=1$ (solid) and
$\kappa=8$ (dashed)(color online). } \label{clustersize}
\end{figure}

There are three key parameters which control the dynamics: 
i) persistence of particle motion, regulated by $F$, 
ii) the packing fraction $\eta$, i.e., the
area occupied by rods divided by the total area ($\eta
= N a/A$, where $N$ is the number of particles in the
system, $a$ is the area of a single particle, and $A$ is the
total area of the box), and iii) the length-to-width aspect ratio
$\kappa$ ($\kappa=L/W$, where $L$ is the length
and $W$ is the width of the rods).
Simulations yield an increase of cluster formation with increasing 
$\kappa$ or $\eta$, see Fig. \ref{fig2}.
Individual clusters are defined by connected particles that have non-zero overlap area. 
Simulations can be characterized by the mean maximum cluster
size, $M$, and the weighted cluster size distribution, $p(m)$, which indicates
the probability of finding a given particle inside a cluster of mass $m$. 
Fig. \ref{clustersize}a shows that for a given $\eta$, $M$ seems to
saturate after the critical $\kappa_{c}$ which is defined as the value 
of $\kappa$ for which the shape of $p(m)$ changes from unimodal to bimodal.
In Fig.
\ref{clustersize}b typical shapes of $p(m)$  are shown: before
clustering and corresponding to low values of $\kappa$ (circles),
and after clustering and corresponding to large values of $\kappa$
(crosses). 
We define the onset of clustering by the
emergence of a second peak in $p(m)$.
We have also tested the robustness of the model against fluctuations
by inserting additive noise terms $R_i/\zeta_i$ in Eqs. (\ref{update_orientation}),
which correspond to a switch from active to active Brownian particles \cite{active-brownian}. 
We found that clustering is still present in rods of the latter kind, albeit the transition
is moved to larger values of $\kappa$ and $\eta$.  
Clustering was absent in all simulations with purely Brownian rods ($F=0$).

%
{\it Mean field approximation (MFA).---} We have studied the clustering
effects described above through a MFA by deriving kinetic equations
for the number $n_j$ of clusters of a given size $j$. 
The equations for $n_j$ contain terms for cluster fusion and fission. 
For the fusion terms we have adopted kinetic equations originally derived 
for coagulation of colloids \cite{chandrasekhar}, while the fission
terms are empirically defined from the typical behavior seen in the
above simulations.
The numbers $n_j$ change in time - we have 
$\left\{
n_{j}\left(t\right)\right\} _{j=1}^{\infty}$, where
$n_{j}\left(t\right)$ is the number of clusters of mass $j$ at time
$t$.

This description neglects the geometry of clusters as well as spatial 
fluctuations. 
This allows us
to consider a single rate constant for all possible collision
processes between clusters of mass $i$ and $j$, as well as a unique
disintegration constant for any cluster of mass $i$.
In addition we make four crucial assumptions:
i) The total number of particles in the system,
$N=\sum_{j=1}^{N}jn_{j}\left(t\right)$, is conserved.
ii) Only binary cluster collisions are considered. Collisions
between any two clusters are allowed whenever the sum of the cluster
masses is less or equal to $N$. 
iii) Clusters suffer spontaneous fission only by losing individual 
particles at the boundary one by one, {\it i. e.} a cluster can only decay by a
process by which a
$j$-cluster split into a single particle plus a
$\left(j-1\right)$-cluster. 
This is motivated by observations in the above simulations.
iv) All clusters move at constant speed, $\widetilde{v} \approx F/ \zeta_{\parallel}$, 
which implies that rods in a cluster have high orientational order and interact only 
very weakly with their neighbors. 
Under all these assumptions the evolution of the $n_j$'s is given
by the following $N$ equations:
\begin{eqnarray}
\dot{n}_{1}&=&2B_{2}n_{2}+\sum_{k=3}^{N}B_{k}n_{k}-\sum_{k=1}^{N-1}A_{k,1}n_{k}n_{1}
\nonumber \\
\dot{n}_{j}&=&B_{j+1}n_{j+1}-B_{j}n_{j}-\sum_{k=1}^{N-j}A_{k,j}n_{k}n_{j}
\nonumber \\
&&+\frac{1}{2}\sum_{k=1}^{j-1}A_{k,j-k}n_{k}n_{j-k} \ \quad \mbox{for} \quad j = 2, .....,N-1
\nonumber \\
\dot{n}_{N}&=&-B_{N}n_{N}+\frac{1}{2}\sum_{k=1}^{N-1}A_{k,N-k}n_{k}n_{N-k} \label{rea}
\end{eqnarray}
where the dot denotes time derivative, $B_{j}$ represents the
fission rate of a cluster of mass $j$, defined by $B_{j}=
(\widetilde{v}/R)\sqrt{j}$, and $A_{j,k}$ is the collision rate
between clusters of mass $j$ and $k$, defined by
$A_{j,k}=(\widetilde{v}\sigma_{0}/A)
\left(\sqrt{j}+\sqrt{k}\right)$. 
$\sigma_{0}$ is the
scattering cross section of a single rod. 
$R$ is the only free parameter and indicates the 
characteristic length a rod at the boundary of a cluster moves before it 
is leaving the cluster in a typical fission event.
We assume $R = \alpha L$ taken into account that longer rods will stay attached to 
cluster for a longer time. 
  
\begin{figure}
\centering
\includegraphics[scale=0.35]{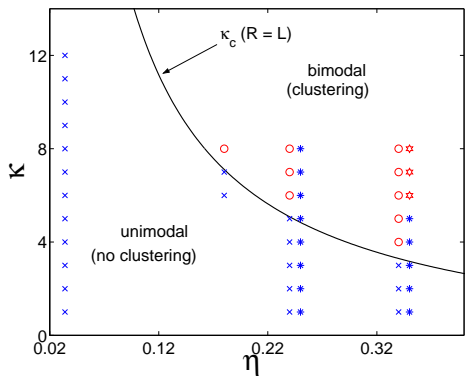}
\caption{$\kappa$-$\eta$ phase diagram. The solid line corresponds
to the transition curve  predicted by equation $\ref{kappa_eta}$.
The symbols indicate IBM simulations ($N=100$). Crosses refer to
unimodal $p(m)$ and circles to bimodal $p(m)$ of active particles.
Stars refer to unimodal $p(m)$ and hexagrams to bimodal $p(m)$
of active-Brownian particles (color online). }
\label{phasediagram}
\end{figure}

Since $\sigma_{0}$ can be approximated by $\sigma_{0}\approx
L+W=\sqrt{a}\left(\sqrt{\kappa}+\frac{1}{\sqrt{\kappa}}\right)$,
the MFA depends only on the parameters $\kappa$, $a$, $A$, 
$\widetilde{v}$ and $\alpha$. 
If one integrates Eqs. (\ref{rea}) with  parameters used in
IBM simulations and an initial condition $n_{j}\left(t=0\right)=N\delta_{1,j}$, 
their solution yields steady state values $n_{j}^0$  for $t \rightarrow \infty$. 
From these values, we obtain a MFA for the weighted cluster size distribution 
 $p(m)=n_{m}^0 m/N$ for given values of the free parameters $R$ resp. $\alpha$. 
The best agreement between the MFA and the IBM simulations is found for a choice of 
$\alpha = 1.0 \pm 0.05$ (see Fig.
\ref{clustersize}b).
Hence, we will use $ R = L$ in the following. 
To understand the relation between the parameters of the model and
clustering effects, we can rescale Eqs. (\ref{rea}) by introducing a new time
variable: $\tau=t \widetilde{v} / \sqrt{a
\kappa}$. 
The resulting equations \cite{dimensionless} depend
only on a dimensionless parameter $P=(\kappa+1)a/A$. 
Note that
$\widetilde{v}\neq 0$ is scaled and does
not affect the qualitative dynamics of the system. 
%
%
In the dimensionless model the parameter $P$ stands for ratio between 
fusion and fission processes and therefore triggers the transition
from a unimodal to a bimodal cluster size distribution.
We can easily establish a transition
criterion, and by using a bisection method, we can accurately
determine the critical transition parameter $P_{c}$. 
Given the system area $A$, the rod area $a$ and the number of rods $N$, 
this method provides a way to calculate $\kappa_{c}$:
\begin{equation} \label{kappa_pc}
\kappa_{c}=P_{c}(N)\frac{A}{a}-1
\end{equation}
At this point it is crucial  that $P_{c}$  depends on
$N$, which is formally  the number of equations in the MFA.
By numerically solving the MFA equations for different particle numbers $N$ up
to $N=1024$, we find  $P_{c}(N)\propto
N^{-1.026 \pm 0.023}$.
This result indicates that in the MFA the critical parameter value $\kappa_c$ for the 
clustering transition does not depend on the number of particles. 
This does not imply that the weigthed distribution $p(m)$ is independent of $N$; 
in fact, we find that the probability for a rod to be in a large cluster increases 
with $N$. 
We proceed by assuming that $P_{c}$ is inversely proportional with
$N$, we can express
 $\kappa_{c}$ as a simple function of the packing fraction:
\begin{equation} \label{kappa_eta}
\kappa_{c} = C/\eta-1
\end{equation}
where the constant is found to be $C \approx 1.46$. 
The $\kappa$-$\eta$
phase diagram (Fig. \ref{phasediagram}) shows a reasonable agreement
of the transition line given by Eq. (\ref{kappa_eta}) and the IBM simulation
results. 
%
So, for the range of parameters used in the IBM, we retrieve in the
MFA the unimodal shape of the weighted cluster size distribution for 
small values of $\kappa$ and $\eta$, and the bimodal shape for large
values of the two parameters. 
Fig.\ref{clustersize}b gives a comparison of the cluster size distribution
in the IBM and MFA.

In summary, we have found non-equilibrium clustering for interacting
self-propelled rod shaped-particles with sufficient packing density
$\eta$ and aspect ratio $\kappa$ in simulations.
The rods interact via strong short range repulsive interactions that
approximate excluded volume interactions.
The onset of clustering has been defined by a transition from a unimodal to bimodal
cluster size distribution.
This transition is reproduced by a mean-field description of the
cluster size distribution, which yielded a simple criterion, $\kappa
= C/\eta - 1$, for the onset of clustering.
This functional form  with $C \approx 1.46 $ provides a good fit to the results of the simulations.
The high density inside the cluster leads also to alignment of rods and
coordinated motion of all particles in the cluster.
The transition to clustering defined here is practically independent
of the system size resp. the number of particles.
It is instructive to compare our result rewritten in the form
 $\kappa\eta + \eta \approx 1.46$ with the formula for the
 isotropic-nematic transition  $\kappa\eta \approx 4.7 $ found in the two-dimensional version
\cite{raveche} of Onsager³s mean-field theory for Brownian rods \cite{onsager}.
This shows that actively moving rods can achieve alignment at much lower densities
than Brownian rods resp. particles in equilibrium systems.
The clustering phenomenon is absent in simulations with isotropic self-propelled
particles as well as with Brownian rods.
Our model provides also an alternative explanation for collective behavior of 
rod-shaped objects - previous swarming models have achieved 
aggregation and clustering 
by assuming attractive long-range interactions 
\cite{theocollectivemotion,active-brownian}.  
With respect to biology, our observation offers a simple physical
explanation for the formation of clusters in many gliding rod-shaped bacteria,
that often precedes the formation of biofilms and the appearance of
more complex patterns. 
%
%
%
%
{\bf Acknowledgement:} We acknowledge financial support of Deutsche
Forschungsgemeinschaft (DFG) through grant DE842/2 and
fruitful discussions with L. Morelli, L. Brusch, J. Starruss and L. Sogaard-Andersen.





\begin{thebibliography}{99}
\bibitem{parameters}
Parameters: $a=0.2$, $1 / \zeta_{\parallel} =0.1$,
$1 /\zeta_{\perp} =0.04$, $1/\zeta_{\theta}=0.5$, $F=0.4$,  $\gamma = 0.16$, $\phi =
1.96$, and $\beta = 1$.  
Additive white noise terms: $ \overline{R_{\parallel}^{2}}/ \zeta_{\parallel}= 
\overline{R_{\perp}^{2}}/ \zeta_{\perp} =\overline{R_{\theta}^{2}}/ \zeta_{\theta}
=0.0833$ for active-Brownian particles. 
%
 Simulations were performed with $N=50$,
$100$ and $200$, and simulation times $10^4$ in discrete steps $\delta t = 0.1$.
\end{thebibliography}
\end{document}